\documentclass[a4paper,noarxiv,twocolumn]{quantumarticle}
\usepackage[utf8]{inputenc}
\usepackage[english]{babel}
\usepackage[T1]{fontenc}
\usepackage[ruled,vlined]{algorithm2e}

\usepackage{graphicx}
\usepackage{amsmath}
\usepackage{bbold}

\usepackage{hyperref}
\usepackage{breakurl}

\title{Efficient quantum gate decomposition via adaptive circuit compression}

\author{Péter Rakyta}
\affiliation{Department of Physics of Complex Systems, E\"otv\"os  Lor\'and  University, Budapest, Hungary}
\affiliation{Wigner Research Center for Physics, 29–33 Konkoly–Thege Miklos Str., H-
1121 Budapest, Hungary}

\author{Zoltán Zimborás}
\affiliation{Wigner Research Center for Physics, 29–33 Konkoly–Thege Miklos Str., H-
1121 Budapest, Hungary}
\affiliation{Algorithmiq Ltd, Kanavakatu 3C 00160 Helsinki, Finland}

\begin{document}

\maketitle

\begin{abstract}
In this work, we report on a novel quantum gate approximation algorithm based on the application of parametric two-qubit gates in the synthesis process. The utilization of these parametric two-qubit gates in the circuit design allows us to transform the discrete combinatorial problem of circuit synthesis into an optimization problem over continuous variables.
The circuit is then compressed by a sequential removal of two-qubit gates from the design, while the remaining building blocks are continuously adapted to the reduced gate structure by iterated learning cycles.
We implemented the developed algorithm in the SQUANDER software package and benchmarked it against several state-of-the-art quantum gate synthesis tools. Our numerical experiments revealed outstanding circuit compression capabilities of our compilation algorithm providing the most optimal gate count in the majority of the addressed quantum circuits.
\end{abstract}

\section{Introduction}

Quantum computers are expected to provide significant speedup compared to the best classical algorithms for a number of applications. Many of the applications are based on the most well-known quantum algorithms such as Shor's integer factorization \cite{doi:10.1137/S0097539795293172} Grover's search \cite{PhysRevLett.79.325} and the Harrow-Hassidim-Lloyd algorithm for solving linear systems of equations \cite{PhysRevLett.103.150502}. However, recently also other promising schemes were developed to exploit quantum resources in solving computational problems such as variational quantum optimization \cite{harrigan2021quantum}, quantum approximate optimization\cite{2014arXiv1411.4028F}, variational quantum eigensolvers \cite{Peruzzo2014,Kandala2017,arute2020hartree}, and quantum simulations of many-body phenomena \cite{Smith2019, satzinger2021realizing}.

Some of these algorithms were already implemented on quantum hardware obtaining at least qualitatively justified results.
From these experiments we have learned that the dedicated optimization of the quantum programs for today's intermediate scaled quantum processors (NISQ) is of high importance, involving to find the most optimal configuration and routing of the hardware resources.
For example, depending on the underlying architecture\cite{Linke3305} the operation error characteristic for one- and two-qubit gates might differ even by an order of magnitude \cite{10.1145/3297858.3304007}. 
In addition, most of the quantum gate based NISQ processors exhibit limited connectivity between the qubits implying unavoidable boundary conditions in quantum gate synthesis.
(An exception are trapped ion based architectures\cite{Debnath2016,Schafer2018}, providing all-to-all connectivity between the qubits.)

In general, quantum algorithms can be described by unitary
transformations and projective measurements acting on the $2^n$-dimensional Hilbert space spanned by the computational basis states of $n$ quantum bits (qubits) involved in the program.
The unitary transformations can be decomposed in terms of elementary unitary transformations (quantum logic gates), supported by the hardware.
A widely used method to characterise the complexity of a quantum circuit (i.e. a quantum program) is to enumerate the number of the elementary gates involved in the circuit\cite{PhysRevA.52.3457} (or alternatively, the depth of the circuit composed by these elementary gates).
Obtaining the most optimal gate structure for a quantum programs implies a discrete combinatorial problem that was addressed in several recent works\cite{Sivarajah_2020,davis2019heuristics,9259942,patel2021robust,nagarajan2021quantumcircuitopt}.
References \cite{madden2021best} and \cite{rakyta2022approaching} showed that the most challenging part of the problem can be bypassed using static gate structure, simplifying the decomposition problem to an optimization over continuous parameters associated with the incorporated single-qubit gates.
These methodologies paved a way to approximate quantum programs with a $CNOT$ gate count approaching the theoretical lower bounds, showed to be sufficient to decompose any unitary $U$ \cite{PhysRevA.69.062321}.
However, in the era of NISQ devices when the computational error accumulated over the subsequent quantum gate operations plays an important aspect, it is highly desirable to find an optimal decomposition (or an accurate approximation) of a unitary consisting of gate count as few as possible. 

To this end a valuable work of Ref.~\cite{2020arXiv200304462Y} reformulated the combinatorial problem into continuous variable optimization problem. This algorithm attempts to decompose an $n$-qubit unitary in terms of $n/2$-qubit general unitary gates, repeating the procedure until the last two-qubit gates can be decomposed via KAK algorithm\cite{tucci2005introduction}.
References \cite{9259942} and \cite{davis2019heuristics}, on the other hand, proposed a systematic strategy implementing A$^*$ search algorithm\cite{4082128} to comply with the combinatorial search problem. 
Later, this approach was significantly improved by narrowing down the search space utilizing constant prefix solutions in the search for an optimal quantum circuit\cite{smith2021leap}.
The implementation of these algorithms provide the core engines working within the quantum syntheses software tools QFAST\cite{qfast_github} and QSearch\cite{qsearch_github}.

In this work we present a novel quantum gate approximation algorithm implemented in the SQUANDER\cite{SQUANDER_github} package.
Our methodology is based on iterations of \emph{adaptive circuit compression} in which an initial quantum circuit is sequentially compressed by the removal of controlled two-qubit gates from the circuit. 
In addition, we compare our implementation included in the SQUANDER package with the state of the art syntheses tools freely accessible on the Internet. 
Motivated by recent benchmark comparison published in Refs.~\cite{2020arXiv200304462Y,smith2021leap} we chose to compare SQUANDER with QFAST and QSearch (extended by the LEAP\cite{smith2021leap} extension) synthesis tools.
We also included the QISKIT\cite{qiskit_org} package using its \emph{transpile} utility in our comparison since it is one of the most frequently used quantum compiler tool on the field.
In the benchmark we tested the decomposition of $3$, $4$ and $5$-qubit unitaries from online database \cite{ibm_mapping} containing series of circuits published as part of the Qiskit Developer Challenge, a public competition to design a better routing algorithm.
Our tests also included known circuits such as mul, add, QFT\cite{10.1093/imamat/25.3.241}, HLF\cite{doi:10.1126/science.aar3106}, and algorithms like variational quantum eigensolver (VQE) \cite{McClean_2016} circuits and Transverse Field Ising Model (TFIM)\cite{Shin2018,PhysRevB.101.184305} simulating the time evolution of an Ising system. 

In great majority of the addressed quantum programs SQUANDER provided significantly lower gate count compared to other synthesis tools involved in our numerical experiments. 
For example, for a set of unitaries reported in Table \ref{table:IBM_CNOT_gates} we achieved more than $50\%$ circuit compression in $21\%$ percent of the use cases while keeping the circuit fidelity close to unity.
Moreover, in $68\%$ of the examples the compression achieved by the SQUANDER package exceeded $10\%$.

The rest of the paper is organised as follows: in Secs.~\ref{sec:costfunction} and \ref{sec:alg} we describe the theoretical background working behind the adaptive circuit compression iterations.
Then we provide our numerical results on benchmarking the SQUANDER package with other synthesis tools.
In Sec.~\ref{sec:benchmark}, we report the benchmark result for all-to-all qubit connectivity topology; while Secs.~\ref{sec:architecture} and \ref{sec:circuit_optimization} are dedicated to discussing the circuit synthesis on linear connectivity architecture and optimisation of deep quantum circuits, respectively.
In Sec.~\ref{sec:parametrized}, we outline a numerical approach to significantly reduce the synthesis time of parametric quantum circuits paving the way to utilize optimization based circuit systhesis strategies in variational quantum algorithms.
Finally, we conclude our work in Sec.~\ref{sec:conclusion}.
The most optimal quantum circuit decompositions obtained by the adaptive circuit compression algorithm are provided within the SQUANDER package in QASM format accessible via a GitHub repository\cite{SQUANDER_github}.

\section{Fundamental quantities characterizing the quality of a gate synthesis} \label{sec:costfunction}

The numerical characterization of approximate gate synthesis and optimisation was addressed by several previous projects. 
In order to quantify the 'distance' of the synthesized $d\times d$ unitary $V$ from the original unitary $U$ Ref.~\cite{Khatri2019quantumassisted}, for example, introduced the Hilbert-Schmidt test
\begin{equation}
    C_{HST}(U,V) = 1 - \frac{1}{d^2}\left|\textrm{Tr}\left(V^{\dagger}U\right)\right|^2\,. \label{eq:hilbert-schmidt}
\end{equation}
The gate fidelity $\overline{F}(U,V)$, measuring the 'closeness' of two unitaries $U$ and $V$, is obtained
by averaging the state fidelities of output states (after the $U$ and $V$ evolution, respectively) over the Haar distribution  \cite{Khatri2019quantumassisted}; this can be calculated from the Hilbert-Schmidt test using the equation
\begin{equation}
    \overline{F}(U,V) = 1 - \frac{d}{d+1}C_{HST}(U,V)\;.
\end{equation}
Madden et al. \cite{madden2021best}, on the other hand, used a different, Frobenius norm based metric to quantify the distance between the two unitaries $U$ and $V$:
\begin{equation}
    f(U,V) = \frac{1}{2}\left\|V-U \right\|_F^2 = d - \textrm{Re}\left[{\textrm{Tr}}(U^{\dagger}V)\right], \label{eq:frobenius}
\end{equation}
and defined a Frobenius based fidelity $\overline{F}_F(U,V)$ by
\begin{equation}
    \overline{F}_F(U,V) = 1 - \frac{d}{d+1} + \frac{1}{d(d+1)}\left( d - f(U,V)\right)^2 \label{eq:frobenius_fid}
\end{equation}
It can be shown that in general $\overline{F}_F(U,V)\leq \overline{F}(U,V)$ holds on\cite{madden2021best}.

Both of Eqs.~(\ref{eq:hilbert-schmidt}) and (\ref{eq:frobenius}) can be efficiently used as a cost function in optimization problems formulated to find the best approximation of the unitary $U$, since they are relatively easy to evaluate numerically.
In addition, one can also derive analytical expressions for their gradient components with respect to the free parameters of the decomposing quantum circuit improving the numerical efficiency of gradient descent optimization calculations, such as the Broyden–Fletcher–Goldfarb–Shanno (BFGS) optimization algorithm\cite{kelley1999iterative} used in the SQUANDER package.
Due to the lower numerical complexity of Eq.~(\ref{eq:frobenius}), however, in this work we used Eq.~(\ref{eq:frobenius}) for the cost function in the numerical optimizations.
\begin{figure}
     \centering
     \includegraphics[width=0.4\textwidth]{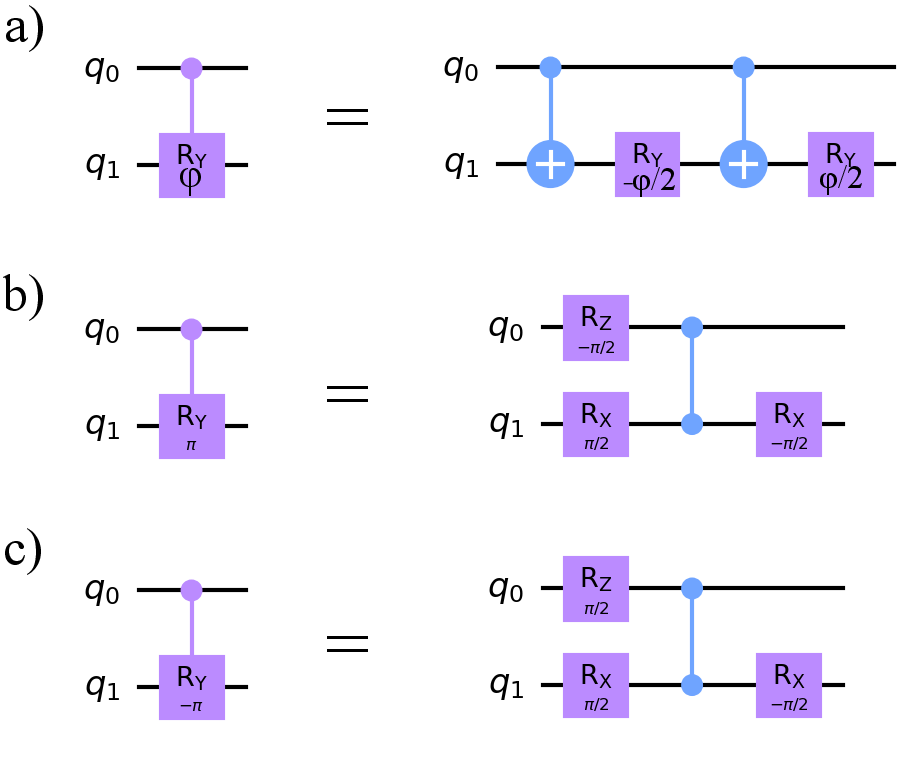}
     \caption{Mapping the controlled $R_y$ gate to single- and constant two-qubit gates. a) In general case, the controlled $R_y$ rotation can be expressed in terms of two $CNOT$ and two $R_y$ gates. b) and c) shows the expansion of the controlled $R_y$ gate for special parameter values $\theta=\pm\pi$, when it can be expressed in terms of singe two-qubit gate.}
     \label{fig:cry_gates}
 \end{figure}

\section{Description of the adaptive decomposing algorithm} \label{sec:alg}

As we mentioned in the introduction, our methodology is based on iterations of \emph{adaptive circuit compression} in which an initial quantum circuit is sequentially compressed by the removal of parametric two-qubit gates from the circuit. 
This way the quantum circuit gets compressed, until no further two-qubit elements can be removed from the design. 
In contrast with the compression strategy of Ref.~\cite{madden2021best}, in our approach the two-qubit gates used in the synthesis process are controlled rotation ($CR$) two-qubit gates, tunable via a continuous parameter. 
The advantage of using such parametric two-qubit gates lies in their versatile ability to express quantum circuit elements.
At some specific parameter values $CR$ gates can be considered as trivial, non-entangling gates, while at some other parameter values they can be mapped to special two-qubit gates such as controlled not ($CNOT$) or controlled Z ($CZ$) gate.
In general (when the parameter value is different from the previously mentioned special cases) a $CR$ gate can be decomposed in terms of two $CNOT$ gates.
Since a $CNOT$ gate can be considered as a special case of the $CR$ gate (up to single-qubit transformations), any $CNOT$ gate in a quantum circuit can be replaced by a $CR$ gate.
Consequently, the set of the $CR$ and the general single qubit rotation ($U3$) gates is universal, capable to synthesise any unitary $U$.
In our specific implementation we used controlled $R_y$ rotation gates (i.e. rotations around axis $y$) which can be decomposed into elementary gates according to Fig.~\ref{fig:cry_gates}. 
This way we can reformulate the structural combinatorial problem of placing the elementary two-qubit gates in a circuit into an optimization problem over continuous variables.

In this context the terminology 'adaptive circuit compression' means that $CR$ gates becomes correlated during the compression, all of the $CR$ gates react upon the removal of a two-qubit block from the design, and our circuit compression approach would not be limited to local two-qubit gate cancellations.
If the optimization problem associated with the reduced gate structure can be solved, then the chosen two-qubit block turns to be a redundant one and can be removed from the system.
Some of the $CR$ gates might become trivial during the optimisation process being also removable from the system in a single compression cycle.
\begin{figure}
     \centering
     \includegraphics[width=0.5\textwidth]{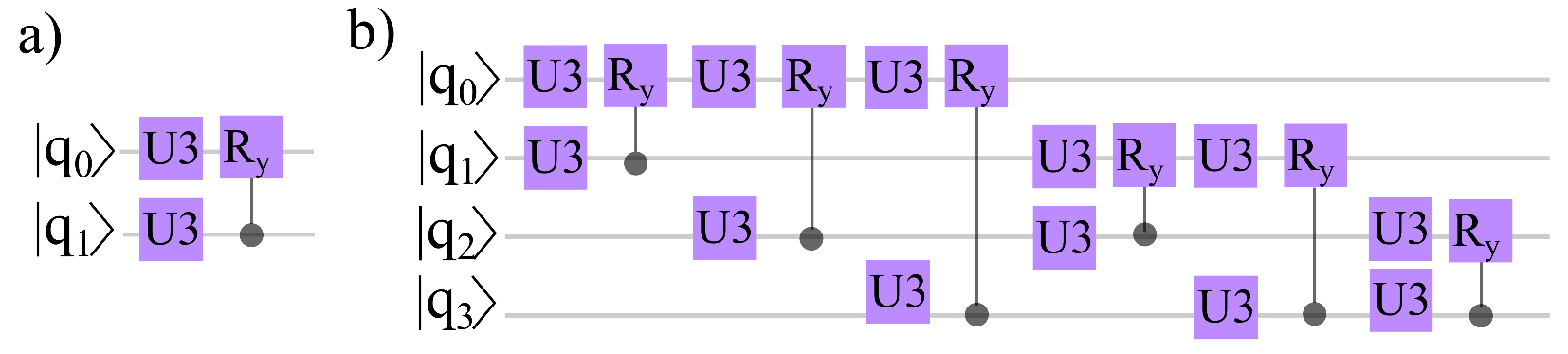} 
     \caption{a) The two-qubit building block used in the adaptive circuit compression algorithm consisting of two $U3$ rotation gates acting on the individual qubits and a controlled $R_y$ rotation gate.
     b) A 4-qubit unit cell of the two-qubit building blocks from which the initial quantum circuit approximating the unitary $U$ is constructed. (see the main text for further explanation.)}
     \label{fig:building_blocks}
 \end{figure}

In order to find a decomposition of the unitary $U$ with optimised gate count, we first need to construct an initial quantum circuit approximating $U$. 
Motivated by recently reported success of using pre-designed gate structures\cite{madden2021best,rakyta2022approaching} in unitary decomposition we construct the initial decomposing circuit from periodically repeated unit cells made of two-qubit building blocks.
Each of the two-qubit building blocks of the unit cell is constructed from two $U3$ single-qubit rotations acting on each of the qubits of the building block, and a single controlled $R_y$ gate as shown in Fig.~\ref{fig:building_blocks}.a).

On fully connected architecture (when there is no limitation in qubit to qubit connections), the unit cell would contain each combination of the qubit-pairs.
Thus, in the case of $N$ qubits, the unit cell of the structure, referred as \emph{sequ} structure in Ref.~\cite{madden2021best}, would contain $N(N-1)/2$ two-qubit building blocks as it is shown in Fig.~\ref{fig:building_blocks}.b) for a $4$-qubit case.
The initial gate structure (being the subject of forthcoming compression cycles) is constructed from these unit cells. During the initial optimization phase the lowest number of the unit cells is determined by trying to solve the optimization problem while sequentially increasing the circuit depth, i.e. after each unsuccessful optimization iteration one unit cell is added to the design.  

After an initial quantum circuit approximating the unitary $U$ was successfully constructed, the algorithm proceeds with compression cycles.
In each iterations we randomly select one of the two-qubit building blocks in the circuit and try to remove it from the system. 
The building block is removable from the system if the remaining circuit elements can be adopted to the changed structure by finding a new solution for the optimization problem.
Finally, when the algorithm does not find more removable two-qubit blocks in the design, we expand the $CR$ gates in terms of elementary two-qubit gates according to Fig.~\ref{fig:cry_gates} and finish the gate synthesis of the unitary with a final optimization iteration.

\section{Benchmark comparison, numerical analysis} \label{sec:benchmark}

As mentioned in the introduction, we benchmarked the adaptive compression algorithm developed in this work with the most efficient decomposing tools reported in Refs.~\cite{2020arXiv200304462Y} and \cite{smith2021leap}.
On one hand, \emph{QFAST} showed outstanding computational performance and robust success rate in the decomposition of $3$, $4$ and $5$-qubit gates addressed in our numerical experiments.
The numerical efficiency of the tool originates from the strategy to utilize parametrized generic unitaries in the synthesis algorithm.
Due to the generic nature of the building blocks, however, there is no straightforward way to keep low the gate count in the synthesised quantum circuits. 
The \emph{QSearch} tool, on the other hand, is designed to find the most optimal decomposition of a unitary via A$^*$ tree search. 
However, due to the nature of the algorithm, the execution time of the gate synthesis rapidly increases with the number of possible trial circuits.
Recently the scaling of the search space was notably moderated by the LEAP extension\cite{smith2021leap} significantly reducing the execution time of gate synthesis.
Though the narrowed search space may drive the algorithm to avoid the most ideal decomposition, recent benchmark calculations showed\cite{smith2021leap} quite decent decomposing capabilities, for some special quantum gates even the ideal $CNOT$ gate count was achieved. 
\begin{figure}
     \centering
     \includegraphics[width=0.4\textwidth]{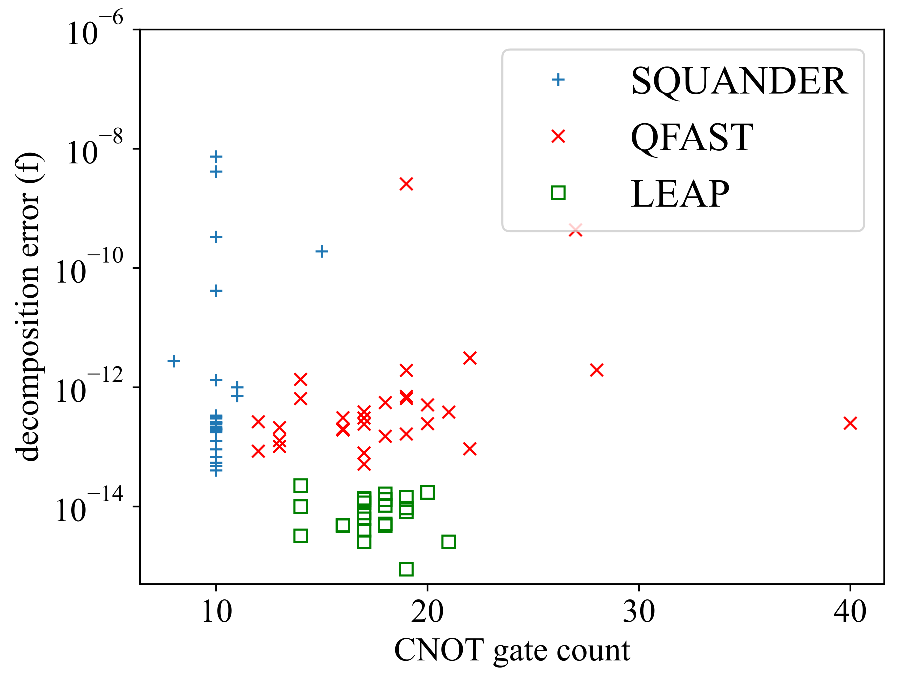}
     \caption{Decomposition results of the $4$-qubit unitary labeled by \emph{adder-q4} in the benchmark examples of the software QFAST.
     The colored crosses represents the final error $f$ defined by Eq.~(\ref{eq:frobenius}) of the optimization as a function of the number of the $CNOT$ gates incorporated in the individual quantum circuits approximating the unitary $U$. The different colors stand for the results obtained by the QFAST, QSearch (LEAP) and the SQUANDER packages.}
     \label{fig:adder-q4}
 \end{figure}
In our benchmark comparison we examined the capabilities of the chosen synthesis tools on statistical basis.
Since optimization algorithms give different results from run to run, we repeated the unitary synthesis multiple times for each unitary and for each software tool.
During the benchmark Qiskit, QFAST and SQUANDER showed a stable execution during the synthesis, we executed them between $10$--$30$ times for each unitary depending on the number of the qubits and $CNOT$ gates.
Then we chose the most optimal decompositions incorporating the fewest $CNOT$ gate count while reaching decomposition fidelity close to unity.
The execution of the QSearch package (using it's LEAP solver), on the other hand, turned to be more irregular.
During our work we repeatedly experienced a situation when the first few runs of the decomposition of a given unitary took only some minutes, while the next turn of the iterations did not finish in the $12$h time limit that we set as a hard limit in our numerical experiments.
We also experienced abrupt program terminations due to memory errors.
In such cases we employed only the first few iterations in our statistics.

We performed the benchmark calculations on a computing server equipped with $32$-Core AMD EPYC 7542 Processor (providing $64$ threads with multi-threading) and with $128$GB of memory. 
Figure \ref{fig:adder-q4} demonstrates a data sets obtained for the decomposition of the $4$-qubit unitary labeled by \emph{adder-q4} taken from the benchmark examples of the QFAST package.
The figure shows the results of $30-30-30$ runs of the optimization based synthesis tools involved in our benchmark. 
As one can see, the results corresponding to the individual software tools (labeled by different colors) are in general close to each other, forming obliterated clusters of data points.
The QSearch package provided the most accurate quantum circuits approximating the initial unitary.
This observation was typical during our benchmark, QSearch turned to be the most precise synthesis tool in our comparisons.
The decomposing precision of the QFAST and SQUANDER packages are close to each other, providing somewhat less precise circuits, but still close to the numerical precision of QSearch.
From the point of $CNOT$ gate count SQUANDER systematically provided the most optimal quantum circuits, with the minimum of $8$ $CNOT$ gates in the most optimal decomposition.    

In Tables \ref{table:IBM_CNOT_gates} and \ref{table:QFAST_CNOT_gates} we summarize the results of further decomposition experiments. 
\begin{figure*}
    \centering
    \includegraphics[width=\textwidth]{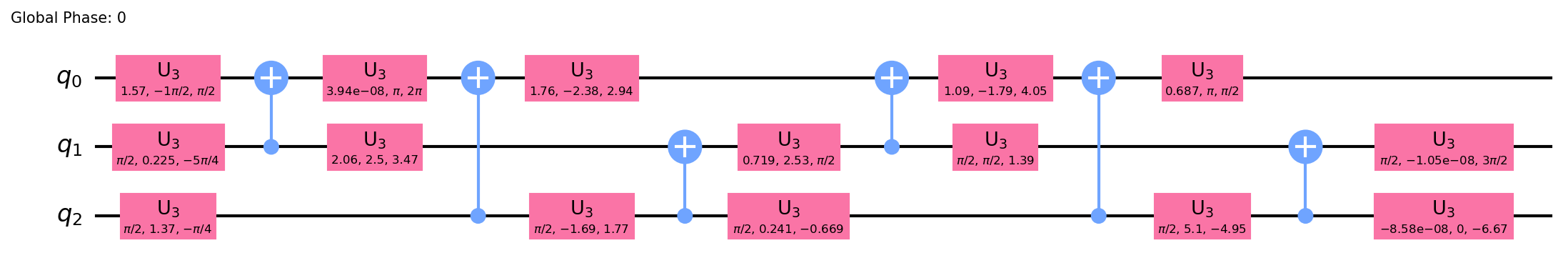}
    \caption{$6$ $CNOT$ approximation of the $3$-qubuit gate $ham3\_102$ from online database \cite{ibm_mapping} obtained by the adaptive decomposition algorithm implemented in the SQUANDER package. The error of the approximation is $f=4\times10^{-12}$.}
    \label{fig:ham102_6CNOT}
\end{figure*}
Namely, Table \ref{table:IBM_CNOT_gates} shows the decomposition results for unitaries taken from the online database of \cite{ibm_mapping}.
From the database we have chosen various $3$, $4$ and $5$-qubit unitaries and tried to decompose them using the above mentioned synthesis tools.
In Table \ref{table:IBM_CNOT_gates} we also provide the $CNOT$ gate count of the quantum circuits synthesized with QISKIT and the initial $CNOT$ gate count of the circuits imported from the QASM files of the database \cite{ibm_mapping}.    
\begin{table*}[ht!]
\centering
\begin{tabular}{||c | c | c | c | c |  c | c | c  | c||} 
 \hline
  File name  & Initial & QISKIT & \multicolumn{2}{c|}{SQUANDER\cite{SQUANDER_github}} &  \multicolumn{2}{c|}{QFAST\cite{2020arXiv200304462Y}} & \multicolumn{2}{c||}{QSEARCH\cite{smith2021leap}} \\
      & $CNOT$ & $CNOT$ & $CNOT$ & $\overline{T}\,[s]$ & $CNOT$  & $\overline{T}\,[s]$ & $CNOT$  & $\overline{T}\,[s]$ \\ [0.5ex] 
 \hline\hline
 4gt5\_77 & $58$ & $338$ & $19$ & $2855$ & $24$ & $222$ & - &  -   \\ 
 4gt13\_91 & $49$ & $187$ & $23$ &  $1296$ & $25$ & $732$ & $48$ &  $>2324^{\ref{leaptime}}$   \\ 
 ham3\_102 & $11$ & $15$ & $6$ &  $4.9$ & $7$ & $3.2$ & $8$ & $2.6$  \\ 
 4gt5\_76 & $46$ & $529$ & $24$ & $1711$ & $29$ & $476$ & -  & -  \\ 
 alu-v0\_26 & $38$ & $204$ & $23$ &  $7900$ & $42$ & $912$ & $29$ & $9284$  \\ 
miller\_11 & $23$ & $18$ & $8$ &  $7$ & $9$ & $5.4$ & $10$ & $4.5$  \\ 
rd32\_v1\_68 & $16$ & $66$ & $9$ &  $23.9$ & $13$ & $21.6$ & $13$ & $615$  \\  
one-two-three &  &  &  &  &  &  &  &  \\ 
-v2\_100 & $32$ & $502$ & $37$ & $5141$ & $52$ & $4142$ & $43$ & $4353$ \\   
4mod5-v0\_20 & $10$ & $526$ & $9$ & $3650$ & $17$ & $166$ & $16$  & $14508$  \\  
alu-v0\_27 & $17$ & $212$ & $17$ & $3452$ & $30$ & $674$ & $34$ & $>3801^{\ref{leaptime}}$  \\   
mod5mils\_65 & $16$ & $73$ & $12$ &  $11162$ & $20$ & $405$ & -  & -  \\ 
ex-1\_166 & $9$ & $20$ & $9$ & $4.4$ & $8$ & $4.7$ & $8$  & $5.9$  \\
decod24-v1\_41 & $38$ & $130$ & $20$ &  $2414$ & $36$ & $413$ & $24$ & $349$  \\ 
alu-v3\_34 & $24$ & $237$ & $25$ & $6090$ & $37$ & $1814$ & $27$  & $7834$  \\  
3\_17\_13 & $17$ & $23$ & $7$ &  $6.5$ & $9$ & $4.2$ & $9$ & $4.3$  \\ 
4gt11\_84 & $9$ & $163$ & $9$ & $642$ & $20$ & $318$ & - & -  \\    
decod24-v0\_38 & $23$ & $48$ & $14$ & $62$ & $23$ & $58$ & $15$ & $285$  \\ 
4mod5-v0\_19 & $16$ & $75$ & $13$ & $701$ & $21$ & $375$ & - & -  \\
4mod5-v1\_22 & $11$ & $168$ & $9$ & $962$ & $13$ & $52$ & $17$ & $82$  \\
alu-v1\_29 & $17$ & $240$ & $19$ & $3820$ & $33$ & $801$ & - & -  \\ 
alu-v1\_28 & $18$ & $331$ & $19$ & $2488$ & $36$ & $607$ & -  & -  \\
4mod5-v1\_23 & $32$ & $74$ & $13$ & $946$ & $40$ & $702$ & -  & -  \\  
4mod5-v0\_18 & $31$ & $671$ & $15$ & $1134$ & $31$ & $266$ & -  & -  \\  
rd32\_270 & $36$ & $522$ & $14$ & $893$ & $27$ & $627$ & -  & -  \\    
rd32-v0\_66 & $16$ & $66$ & $10$ & $29$ & $16$ & $25$ & $13$ & $443$  \\
alu-v3\_35 & $18$ & $249$ & $20$ & $3655$ & $31$ & $1050$ & -  & -  \\  
4gt13-v1\_93 & $30$ & $218$ & $23$ & $2408$ & $38$ & $466$ & $33$  & $21315$  \\  
4mod5-v1\_24 & $16$ & $241$ & $14$ & $5081$ & $33$ & $210$ & $52$ & $>3968^{\ref{leaptime}}$  \\
mod5d1\_63 & $13$ & $76$ & $13$ & $867$ & $29$ & $304$ & - & -  \\  
alu-v4\_36 & $51$ & $193$ & $40$ & $11090$ & $49$ & $2343$ & - & -  \\
4gt11\_82 & $18$ & $419$ & $15$ & $883$ & $22$ & $698$ & $19$ &  $1003$  \\
4gt5\_75 & $38$ & $259$ & $25$ & $7002$ & $37$ & $429$ & $49$ &  $33246$  \\ 
alu-v2\_33 & $17$ & $358$ & $17$ & $2339$ & $31$ & $665$ & $23$  & $>6520^{\ref{leaptime}}$  \\
4gt11\_83 & $14$ & $151$ & $13$ & $1994$ & $15$ & $98$ & $19$ & $1107$  \\    
decod24-v2\_43 & $22$ & $46$ & $9$ & $93$ & $19$ & $44$ & $17$ & $1390$ \\    
4gt13\_92 & $30$ & $161$ & $24$ & $1767$ & $46$ & $1830$ & -  & -  \\ 
alu-v4\_37 & $18$ & $276$ & $18$ & $3509$ & $37$ & $837$ & $32$ & $>2142^{\ref{leaptime}}$  \\  
mod5d2\_64 & $25$ & $129$ & $14$ & $846$ & $26$ & $104$ & $16$ & $>256$\footnote{\label{leaptime}The benchmark comparison was interrupted due to exceeding the time limit of a single decomposition run.}  \\    
 \hline
\end{tabular}
\caption{$CNOT$ gate count comparison of $3$, $4$ and $5$-qubit circuits obtained by decomposing unitaries taken from the online database \cite{ibm_mapping}. The average execution time $\overline{T}$ of the optimization based decomposing tools was measured by averaging at least $10$ individual runs for the SQUANDER and the QFAST packages, while at least $5$ runs for the LEAP package, unless not stated differently. The final error $f$ of the approximation calculated via Eq.~(\ref{eq:frobenius}) was less than $10^{-8}$ in each of the reported decompositions corresponding to Fidelity $\overline{F}_F$ close to unity by an error of $10^{-9}$. The time limit for a single decomposition run was set to $12$h.}
\label{table:IBM_CNOT_gates}
\end{table*}
The table shows the most optimal decomposition results obtained by the individual synthesis tools having error less than or equal to $f=10^{-8}$.
Our results indicate that SQUANDER provides shorter quantum circuits than the other synthesis tools used in our benchmark (except the unitary $ex-1\_166$, for which SQUANDER gave longer circuit by a single $CNOT$ gate).
Moreover, in $21\%$ percent of the unitary decomposition experiments SQUANDER could decrease the number of the $CNOT$ gates by more than $50\%$ compared to the initial quantum circuits imported from the database\cite{ibm_mapping}.
In $68\%$ of the examples the compression was more than $10\%$, showing notably higher compression rate than was reported in Ref.~\cite{madden2021best}.
An example of the synthesised quantum circuit of unitary $ham3\_102$ consisting of $6$ $CNOT$ gates is shown in Fig.~\ref{fig:ham102_6CNOT}.
(The most optimal, SQUANDER synthesised quantum circuits approximating the unitaries in Table \ref{table:IBM_CNOT_gates} can be found on the Github repository of the SQUANDER package\cite{SQUANDER_github} in QASM format.)

The average execution time of the SQUANDER package significantly exceeds QFAST in most cases, being similar to the average time of successful decompositions of the QSearch package. 
The execution time of the QISKIT package took only several second, thus we omit these data from the table.
The increased execution time of the adaptive decomposing algorithm can be explained by the sequentially repeated iterations of the circuit compression process, similarly to QSearch traversing a decomposition tree and expanding the synthesised quantum circuit from iteration to iteration.
Though, the success rate of the SQUANDER package turned to be much reliable than QSearch:
roughly in half of our numerical experiments we did not received any result from the QSearch package in the $12$h time limit. 
These cases are labeled by dashes in Table \ref{table:IBM_CNOT_gates}. 
Unfortunately, the large scale of our numerical experiments did not allow us to give more attempts to relaunch these unsuccessful gate synthesis attempts. 

A set of further decomposition experiments is reported in Table \ref{table:QFAST_CNOT_gates} summarizing synthesis results for unitaries provided in the QFAST package including algorithms such as mul, add, QFT\cite{10.1093/imamat/25.3.241}, HLF\cite{doi:10.1126/science.aar3106}, VQE circuits and TFIM \cite{Shin2018,PhysRevB.101.184305} algorithms. 
\begin{table*}[ht!]
\centering
\begin{tabular}{|| c | c | c | c |  c | c | c | c ||} 
 \hline
  Circuit name  &  QISKIT & \multicolumn{2}{c|}{SQUANDER\cite{SQUANDER_github}} &  \multicolumn{2}{c|}{QFAST\cite{2020arXiv200304462Y}} & \multicolumn{2}{c||}{QSEARCH\cite{smith2021leap}} \\
       & $CNOT$ & $CNOT$ & $\overline{T}\,[s]$ & $CNOT$ & $\overline{T}\,[s]$ & $CNOT$ & $\overline{T}\,[s]$ \\ [0.5ex] 
 \hline\hline
 adder\_q4 & $66$ & $8$ & $18$ & $12$ & $27$ & $14$ & $49$  \\ 
 qft4 & $126$ & $14$ & $117$ & $21$ & $51$ & $15$ & $94$   \\ 
 tfim-4-22 & $218$ & $11$\footnote{In the best decomposition $f=0.0008$ was achieved resulting in fidelity $\overline{F}_F$ close to unity with error of $10^{-4}$.} & $108$ & $14$\footnote{In the best decomposition $f=0.008$ was achieved resulting in fidelity $\overline{F}_F$ close to unity with error of $10^{-3}$.} & $20$ & $12$ & $10$   \\ 
 tfim-4-60 & $218$ & $18$ & $173$ & $14$ & $20$ & $12$ & $6$    \\ 
 tfim-4-80 & $218$ & $18$ & $178$ & $12$ & $12$ & $12$ & $11$    \\ 
 \hline
 hlf\_q5 & $870$ & $8$ & $270$ & $11$ & $119$ & $12$ & $272$   \\ 
 grover\_5Qs011i2 & $570$ & $33$ &$7339$ & $44$ & $3115$ & - & -  \\ 
 mul\_q5 & $77$ & $12$ & $283$ & $18$ & $202$ & $13$ & $281$   \\ 
 qaoa\_q5-1 & $750$ & $20$ & $307$ & $30$ & $807$ & $22$ & $>13579$\footnote{\label{leaptime2}The benchmark comparison was interrupted due to exceeding the time limit.}   \\ 
 qft5 & $582$ & $22$ & $979$ & $38$ & $1346$ & $26$ & $12721$   \\ 
 vqe & $566$ & $13$ & $8048$ & $22$ & $382$ & $17$ &$1184$   \\ 
 \hline
\end{tabular}
\caption{$CNOT$ gate count comparison of $3$, $4$ and $5$-qubit circuits obtained by decomposing  unitaries shipped within the package QFAST. The average execution time $\overline{T}$ of the optimization based decomposing tools was measured by averaging at least $10$ individual runs for the SQUANDER and the QFAST packages, while at least $5$ runs for the LEAP package, unless stated differently. The final error $f$ of the approximation calculated via Eq.~(\ref{eq:frobenius}) was less than $10^{-8}$ in each of the reported decompositions corresponding to Fidelity $\overline{F}_F$ close to unity by an error of $10^{-9}$. The time limit for a single decomposition run was set to $12$h.}
\label{table:QFAST_CNOT_gates}
\end{table*}
From the reported examples we can conclude that circuit synthesis of the benchmarked packages overestimates the gate count of the well known circuits like the $4$ and $5$-qubit QFT algorithms being optimally expressed by $12$ and $20$ $CNOT$ gates, respectively.
Though, among the compared tools SQUANDER package needed only $2$ extra $CNOT$ gates to synthesise these unitaries, while the QFAST and the QSearch packages needed notably higher gate count to achieve similar accuracy. 
Except the unitaries related to the $tfim$ algorithm, SQUANDER turned to give significantly shorter quantum circuits in other examples as well.
In the decomposition of the $tfim$ unitaries the SQUANDER packages seems to be less efficient than the QFAST and QSearch packages. 

During our benchmark analysis we found that the success rate of the SQUANDER package to find an accurate gate decomposition significantly decreases if the input unitary needs more than $30$ $CNOT$ gates for the decomposition.
According to our numerical experiences, whenever the $CNOT$ gate count required to decompose a unitary is larger than $\sim30$, some of the synthesis executions fails to decompose the unitary, i.e. the synthesis finishes with an unacceptable $f\sim\mathcal{O}(10^{-1})$ optimization error.
For example, in the specific case of unitary \emph{one-two-three-v2\_100} approximated by $37$ $CNOT$ gates (see Table \ref{table:IBM_CNOT_gates}.) only $3$ executions from $30$ attempts have successfully finished the synthesis.
The bottleneck of the adaptive decomposition algorithm lies in the construction of the initial gate structure being reduced over the adaptive compression iterations.
In case of deeper circuits it becomes challenging to find a proper solution for the optimization problem during the generation of the initial gate structure, which can be explained by the presence of the redundant two-qubit building blocks (aimed to be removed during the compression cycles) making the optimization problem significantly over-parametrized.
In Sec.~\ref{sec:circuit_optimization} we return to this problem providing a workaround to overcome this issue.

\section{Limited inter-qubit connectivity} \label{sec:architecture}

The ability to find an optimal compilation of quantum programs is undoubtedly a major component of quantum computing.
The ability to compile quantum programs with optimized gate count having in mind the hardware constraints is even more important.
Limitations in the connectivity between the qubits is a constraint typical for todays NISQ hardware.
An undoubted advantage of optimisation based algorithms is the straightforward way to adopt them to such requirements by limiting the optimization problem to the available inter-qubit connections.
Similarly to the synthesis tools compared in this work, the algorithm implemented in the SQUANDER package can account for connectivity limitations as well. 
The Python interface of the SQUANDER package provides a straightforward way to limit the inter-qubit connections to selected qubit pairs.
An example script showing the usage of the python interface including the connectivity restrictions is provided within the source code of the SQUANDER package.
The prescribed connectivity limitations are processed during the synthesis of the initial gate structure by including only those two-qubit building blocks in the unit cells which are provided on the input.  
\begin{table*}[ht!]
\centering
\begin{tabular}{|| c | c | c | c |  c | c | c | c ||} 
 \hline
  Circuit name  &  QISKIT & \multicolumn{2}{c|}{SQUANDER\cite{SQUANDER_github}} &  \multicolumn{2}{c|}{QFAST\cite{2020arXiv200304462Y}} & \multicolumn{2}{c||}{QSEARCH\cite{smith2021leap}} \\
       & $CNOT$ & $CNOT$ & $\overline{T}\,[s]$ & $CNOT$ & $\overline{T}\,[s]$ & $CNOT$ & $\overline{T}\,[s]$ \\ [0.5ex] 
 \hline\hline
 4gt13\_91 & $1042$ & $26$ & $4362$ & $55$ & $1045$ & $35$ & $976$  \\
  4gt5\_76 & $2012$ & $26$ & $2881$ & $51$ & $591$ & $27$ & $5747$ \\
  alu-v0\_26 & $951$ & $32$ & $6837$ &  $56$ & $1806$ & - & -  \\ 
  mod5mils\_65 & $1123$ & $18$ &  $1971$ & $45$ & $885$ & $22$  & $\sim2500^{\ref{leaperror},\ref{leapfew}}$ \\ 
  decod24-v1\_41 & $848$ & $18$ & $1640$ & $45$ & $936$ & $19$ & $3382$  \\ 
  3\_17\_13 & $45$ & $9$ &  $11$ & $8$ & $4.45$ & $9$ & $2.65$  \\ 
  decod24-v0\_38 & $164$ & $14$ & $102$ & $30$ & $42$ & $17$ & $6475$  \\ 
  4mod5-v0\_19 & $1181$ & $17$ & $1825$ & $45$ & $888$ & - & -$^{\ref{leaperror}}$  \\ 
  alu-v1\_29 & $1012$ & $23$ & $1784$ & $62$ & $2647$ & $29$ & $\sim6200^{\ref{leaperror}}$  \\
  alu-v1\_28 & $1349$ & $23$ & $3879$ & $62$ & $1618$ & - & -\footnote{\label{leaperror}QSearch circuit synthesis executions were interrupted with memory error. }  \\
  4mod5-v1\_23 & $1138$ & $18$ & $3886$ & - & - & $24$  & $\sim2500^{\ref{leaperror},}$\footnote{\label{leapfew}Only few successful decompositions were obtained.}  \\  
  alu-v3\_35 & $1076$ & $26$ & $4310$ & $62$ & $2570$ & $34$ & $\sim750^{\ref{leaperror}}$  \\
  4gt13-v1\_93 & $1155$ & $26$ & $7641$ & $91$ & $4201$ & $43$  & $\sim2200^{\ref{leaperror},\ref{leapfew}}$  \\  
  4mod5-v1\_24 & $1413$ & $31$ & $9760$ & $81$ & $2736$ & $44$ & $\sim1200^{\ref{leaperror},\ref{leapfew}}$  \\ 
  decod24-v2\_43 & $162$ & $18$ & $185$ & $28$ & $48$ & $21$  & $2274$  \\  
\hline
\end{tabular}
\caption{$CNOT$ gate count comparison of $3$, $4$ and $5$-qubit circuits obtained by decomposing unitaries taken from the online database \cite{ibm_mapping}. In the decomposition we assumed $0-1-2-3-4$ linear connectivity between the qubits. The average execution time $\overline{T}$ of the optimization based decomposing tools was measured by averaging at least $10$ individual runs for the SQUANDER and the QFAST packages, and at least a single successful run for the QSearch package. The final error $f$ of the approximation calculated via Eq.~(\ref{eq:frobenius}) was less than $10^{-8}$ in each decomposition corresponding to Fidelity $\overline{F}_F$ close to unity by an error of $10^{-9}$. The time limit for a single decomposition run was set to $12$h. Dashes label use cases when we did not succeed to decompose the unitary in this time limit.}
\label{table:IBM_CNOT_gates_linear}
\end{table*}
In Table \ref{table:IBM_CNOT_gates_linear} we report our results on decomposition experiments for the case of linear connectivity having two-qubit gates only between adjacent qubits along a line topology.
In these quantum circuit synthesis experiments the SQUANDER package again turned to give the most optimal gate counts among the benchmarked synthesis tools, except the $3$-qubit unitary $3\_17\_13$.
The $CNOT$ gate count in QISKIT synthesised circuits got significantly increased compared to the all-to-all connectivity case, in all of the addressed examples the number of the $CNOT$ gates reached $\mathcal{O}(10^3)$ (except of a single $3$-qubit unitary $3\_17\_13$).
For example, taking the unitary $4mod5-v1\_23$ the QISKIT quantum circuit consists of $15$ times more $CNOT$ gates than in the all-to-all connectivity case. 
In addition, we did not succeed to decompose this unitary in the $12h$ time limit using the QFAST package, although we made several attempts for it.
Revising the numerical results reported in Table \ref{table:IBM_CNOT_gates_linear} we see that the adaptive optimization algorithm implemented in SQUANDER gave notably lower $CNOT$ gate count than QFAST and QSearch in most use cases.
(Unfortunately, in many cases the Qsearch decompositions were terminated with memory error preventing us to do comprehensive statistical benchmark.)
Compared to the full connectivity case reported in Table \ref{table:IBM_CNOT_gates} the gate count in circuits synthesised by the SQUANDER package increased by $34\%$ in average, we observed the maximal growth of $121\%$ in $CNOT$ gates for the unitary \emph{4mod5-v1\_24}. 
In comparison, QFAST circuits synthesised for linear topology became larger by $79\%$ in average, hitting the maximal growth of $145\%$ for unitary \emph{4mod5-v1\_24}.
In overall, the SQUANDER package turned to be highly efficient in gate synthesis for linear connectivity architecture providing more optimal circuit designs than other software tools used in our comparison.
Even if we did not perform exhaustive benchmark analysis for other connectivity topologies, we believe that the adaptive compression algorithm implemented in the SQUANDER package would perform similarly favourable for other topologies as well.

\section{Quantum circuit optimization} \label{sec:circuit_optimization}

Previously in Sec.~\ref{sec:benchmark} we noticed that in certain cases SQUANDER fails to generate a suitable initial gate structure for the adaptive circuit compression.
\begin{table*}[ht!]
\centering
\begin{tabular}{|| c | c | c | c | c |  c | c | c ||} 
 \hline
   &  &  & \multicolumn{2}{c|}{} &  QFAST+ & \multicolumn{2}{c||}{QFAST+QISKIT+} \\
  Circuit name  & Initial &  QISKIT & \multicolumn{2}{c|}{QFAST\cite{2020arXiv200304462Y}} & QISKIT & \multicolumn{2}{c||}{+SQUANDER\cite{SQUANDER_github}} \\
       & $CNOT$ & $CNOT$ & $CNOT$ & $\overline{T}\,[s]$ & $CNOT$  & $CNOT$ & $\overline{T}\,[s]$ \\ [0.5ex] 
 \hline\hline
  4gt10-v1\_81 & $66$ & $372$ & $66$ & $4481$ & $65$ & $39$ & $65737$  \\
  one\_two\_three-v1\_99 & $59$ & $302$ & $80$ & $7472$ & $74$ & $45$ & $80390$  \\
  one\_two\_three-v0\_98 & $65$ & $213$ & $78$ & $7701$ & $78$ & $61$ & $175994$  \\
  4mod7-v1\_96 & $72$ & $150$ & $38$ & $990$ & $35$ & $33$ & $10255$  \\
  aj\_e11\_165 & $69$ & $337$ & $54$ & $598$ & $52$ & $36$ & $15585$  \\
  alu-v2\_32 & $72$ & $469$ & $52$ & $3540$ & $52$ & $41$ & $33820$  \\
\hline
\end{tabular}
\caption{$CNOT$ gate count comparison of deep $4$ and $5$-qubit circuits obtained by decomposing and optimizing unitaries taken from the online database \cite{ibm_mapping}. The individual columns label the name of the circuit, the initial $CNOT$ gate count of the circuit, the $CNOT$ gate count of the quantum circuit synthesised by QISKIT and QFAST, and the $CNOT$ gate count of the quantum circuit optimized by the QISKIT transpile function, and by the SQUANDER package, respectively. The final error $f$ of the approximation calculated via Eq.~(\ref{eq:frobenius}) was less than $10^{-5}$ in each decomposition corresponding to Fidelity $\overline{F}_F$ close to unity by an error of $10^{-6}$.}
\label{table:IBM_CNOT_gates_opt}
\end{table*}
According to our numerical experiences, this situation occurs when the $CNOT$ gate count needed to synthesise a unitary exceeds $\sim30$.
Unfortunately, we can not provide a mathematically well defined property separating the cases when SQUANDER can or can not construct the initial gate structure. 
Instead, during our numerical calculations we observed that the success rate of constructing the initial gate structure continuously decreases as the complexity of the unitary (i.e. the $CNOT$ gate count needed to decompose it) increases.   

In this section we discuss a possible workaround to optimize quantum circuits consisting of more than $30$ $CNOT$ gates via the adaptive gate optimization algorithm.
Since the adaptive circuit compression algorithm reported in this work does not depend on any specific structure of the circuit, it can be applied on arbitrary initial circuit.
To this end the SQUANDER package is equipped with a dedicated routine to import quantum circuits to further optimize it.
In the first step the two-qubit gates in the quantum circuit are transformed into controlled $R_y$ gates using the inverse of the gate expansions b) and c) of Fig.~\ref{fig:cry_gates}.
Then the transformed quantum circuit can be subjected to the adaptive compression process. 
Table \ref{table:IBM_CNOT_gates_opt}. shows our numerical results obtained on optimizing deeper quantum circuits taken from the online database \cite{ibm_mapping}.

First we decomposed the unitaries by QISKIT and the QFAST package assuming all-to-all connectivity between the qubits. 
As wee can see in columns $3$ and $4$ of Table \ref{table:IBM_CNOT_gates_opt}, QISKIT gives significantly deeper circuit than QFAST. 
The execution time of the QISKIT package, similarly to previous benchmark experiments in Sec.~\ref{sec:benchmark}, took only several seconds (not presented in the table), providing much faster decomposition than QFAST.

In some cases the quantum circuit synthesised by QFAST can be further simplified by some $CNOT$ gates using the \emph{transpile} function of the QISKIT package with optimization level set to $3$, the resulting gate counts are reported in the $6$-th column of Table \ref{table:IBM_CNOT_gates_opt}.
We undertake the resulted quantum circuit to further compression using the SQUANDER package. 
The number of the $CNOT$ gates in the compressed quantum circuit are reported in the $7$-th column of Table \ref{table:IBM_CNOT_gates_opt}.
As one can see, the additional circuit compression ability provided by SQUANDER is significant.
Our benchmark results indicate, that the interplay of the SQUANDER package with other synthesis tools (such as QFAST used in our specific benchmark) provides a very efficient equipment to synthesise deep circuits when none of the individual synthesis tools can provide similarly optimal circuit on their own.
A single disadvantage of this procedure lies in the highly increased computational time reported in the last column of Table \ref{table:IBM_CNOT_gates_opt}.

\section{Parametric quantum circuits} \label{sec:parametrized}

From practical point of view, the increased execution time of the SQUANDER package poses a critical issue in exploiting quantum computing algorithms. 
In variational quantum algorithms a near real-time quantum gate synthesis is necessary to proceed with the hybrid quantum-classical optimization algorithm.
In such situation the $\mathcal{O}(10^3)$s execution time exhibited by SQUANDER and by other optimisation based synthesis tools seems to be unapplicable.
In this section we show that despite of our expectations optimisation based quantum gate synthesis might be still a valuable component in the execution of variational quantum algorithms. 
As a use-case example we consider the Unitary Coupled Cluster (UCC) variational quantum circuit ansatz used for molecular simulations\cite{Romero_2018}. 
In this particular case the full VQE circuit is constructed from smaller building blocks called \emph{Pauli exponentials} defined by Eq.~(5) of Ref.~\cite{Cowtan2020AGC} and parametrized with a single scalar parameter. 
An example to synthesise such a Pauli exponential is shown in Fig.~7(c) of Ref.~\cite{Cowtan2020AGC} utilizing $22$ $CNOT$ gates generated by the \emph{set diagonalization} synthesis approach implemented in $t|ket\rangle$\cite{Sivarajah_2020}.
Also, we synthesised a circuit of a unitary of the above mentioned Pauli exponential evaluated at a randomly chosen parameter value using the SQUANDER package.
In Table \ref{table:IBM_CNOT_gates_parametric} we summarize the $CNOT$ and single-qubit gate count of the synthesised circuits for two connectivity topologies.
When we do not assume any limitations in the inter-qubit connectivity (all-to-all connectivity) we can compare the synthesised circuit to that of Ref.~\cite{Cowtan2020AGC}.
From $40$ synthesis attempts the most optimal circuit contained $19$ $CNOT$ and $43$ single-qubit rotation gates, with fidelity very close to unity by an error of  $10^{-13}$. 
Comparing to the circuit of Fig.~7(c) of Ref.~\cite{Cowtan2020AGC} our decomposition provides less $CNOT$ gates, but significantly more single-qubit rotation gates.
In Table \ref{table:IBM_CNOT_gates_parametric}) we also provide the most optimal gate decomposition for a linear connectivity architecture having two-qubit gates only between adjacent qubits along a line topology, as in Sec.~\ref{sec:architecture}.
The most optimal circuit obtained in $30$ decomposing attempts had $26$ $CNOT$ and $57$ single-qubit gates, having a gate count increment of $\sim36\%$ compared to the full connectivity case.
(This growth is in line with the average gate count increment reported in Sec.~\ref{sec:architecture}.)
\begin{table*}[ht!]
\centering
\begin{tabular}{|| c | c | c | c |  c | c ||} 
 \hline
  Connectivity  &   \multicolumn{2}{c|}{Fig.~7(c) of Ref.\cite{Sivarajah_2020}} &  \multicolumn{3}{c||}{SQUANDER\cite{SQUANDER_github}} \\
       &  $CNOT$ & single qubit gates  & $CNOT$ & single qubit gates & $\overline{T}\,[s]$  \\ [0.5ex] 
 \hline\hline
 all-to-all  & $22$ & $23$ & $19$ & $43$ & $4295$   \\
  linear  & - & - & $26$ & $57$ & $9665$   \\
\hline
\end{tabular}
\caption{The number of single and two-qubit gates in the decomposition of a Pauli exponential circuit shown in Fig.7(c) of Ref.~\cite{Cowtan2020AGC} evaluated at a randomly selected parameter value $\alpha=0.6217\pi$. The final error $f$ of the SQUANDER decomposition calculated via Eq.~(\ref{eq:frobenius}) was less than $10^{-12}$ in each of the reported decompositions corresponding to circuit fidelity $\overline{F}_F$ close to unity by an error of $10^{-13}$.}
\label{table:IBM_CNOT_gates_parametric}
\end{table*}
As reported in Table \ref{table:IBM_CNOT_gates_parametric} the synthesis of a single circuit took in average more than $1$h for the SQUANDER package. 
Here we argue that the quantum circuit synthesised for a given parameter $\alpha$ can be reused to synthesise other circuits corresponding to different values of $\alpha$, significantly reducing the synthesis time.
In such synthesis strategy the parameters of the single-qubit rotation gates would be a nontrivial function of $\alpha$, while the gate structure itself would be fixed.

In particular, in our numerical experiments we made use of the most optimal circuits obtained for a randomly chosen parameter value $\alpha=0.6217\pi$ reported in Table \ref{table:IBM_CNOT_gates_parametric}.
Then we constructed an equidistant parameter range of the $\alpha$ parameters for which we determined the parameters of the single qubit rotation gates to approximate the $\alpha$-dependent Pauli exponent for each value of $\alpha$ in the range. 
In each iteration (taking in average $\sim3$ seconds) we reused the optimized parameters of the previous iteration as the initial value of the optimization process, provided that we monotonously increased the value of $\alpha$ one after another.
This way we have constructed a parameter set consisting of $1000$ different $\alpha$ values and we determined the parameters of the single-qubit rotational gates for all of them.
In order to obtain the quantum circuit for an arbitrary value of $\alpha$ we first interpolate between the pre-determined parameters of the single-qubit rotational gates and take the interpolated set of parameters as a starting point in the circuit synthesis for the chosen $\alpha$.
This way we managed to decrease the average time needed to obtain a quantum circuit of the Pauli exponent down to $\sim0.82$ seconds in both the all-to-all and linear connectivity topologies. 
(The average synthesis time was measured over $10000$ independent runs for randomly chosen $\alpha$ parameters. 
In all of the attempts we reached a fidelity $\overline{F}_F$ close to unity with error of $10^{-10}$.)
We provide our computational Python scripts and data sets as a part of the publicly available SQUANDER package \cite{SQUANDER_github}.

According to our results, the time needed to synthesise parametric circuits can be significantly reduced by fixing the gate structure.
Although we can not provide any mathematically rigorous statement supporting our approach, we have successfully used it in the decomposition of other parametric circuits as well, like the VQE circuit fabrics proposed in Ref.~\cite{Anselmetti_2021} or the time evolution unitary of Ref.~\cite{Leontica2021}.
Also, we believe that it is possible to further optimize the gate synthesis time by using an optimization algorithm with lower computational overhead than the BFGS algorithm currently used in SQUANDER, or by allowing higher decomposition error of the synthesised circuit. 
(Since the initial parameter values determined by interpolation are expected to be close to the minimum, the usage of less robust, but faster optimization algorithms might be possible.)
While our aim here was only to show possible application of optimisation based quantum gate sythesis tools in variational quantum algorithms, we leave a more exhaustive study of our approach to decompose parametric quantum circuits for a future work.

\section{Conclusions}  \label{sec:conclusion}

In this work, we formulated a novel algorithm for the approximate quantum gate synthesis based on adaptive circuit compression cycles. 
The developed synthesis strategy relies on the application of parametric control rotation gates enabling one to reformulate the discrete combinatorial problem of gate synthesis to a continuous variable optimization problem.
We believe that the main advantage of using controlled rotation gates originates from their versatile ability to express quantum circuit elements.
During the process of sequential circuit reductions some of the controlled gates are removed from the circuit, while the remaining two-qubit building blocks in the design are adapted to the reduced structure by iterative learning cycles.

Our algorithm was tested on the decomposition problems of $3$, $4$ and $5$-qubit unitaries taken from publicly available online databases.
Our numerical experiments revealed a remarkable efficiency in getting highly optimised gate counts compared to state-of-the-art gate synthesis tools. 
In the majority of the addressed unitaries, the SQUANDER package implementing the adaptive circuit compression algorithm, provided the most optimal gate count in the resulting quantum circuits with fidelity very close to unity.
Regarding the online database of Ref.~\cite{ibm_mapping} SQUANDER could reduce the number of $CNOT$ gates by more than $50\%$ in $21\%$ percent of the decomposed unitaries.
In $68\%$ of the addressed examples, the obtained circuit compression exceeded $10\%$ (see Table \ref{table:IBM_CNOT_gates} for details).
Except of some specific quantum programs (namely the $tfim$ unitaries), SQUANDER 
also provided the most optimal gate decomposition for the $4$ and $5$-qubit benchmark unitaries shipped with the QFAST package (see Table \ref{table:QFAST_CNOT_gates} for details). 
We also tested our algorithm by approximating quantum programs for a limited connectivity topology, where the qubits are connected to each other over a line.
Our decomposition results for this case are summarized in Table \ref{table:IBM_CNOT_gates_linear} showing an even larger difference in the gate counts corresponding to the individual synthesis tools in favour of the SQUANDER package.
In addition, the developed adaptive circuit compression strategy can also be applied to optimize deep quantum circuits generated by other tools. 
This capability turned to be particularly useful in the optimization of quantum programs incorporating more than $\sim30$ $CNOT$ gates when the success rate of the SQUANDER package to synthesise the initial quantum circuit approximating the unitary becomes low.
In such situation the combination of the SQUANDER package with other synthesis tools provides quite an efficient synthesis strategy (see Table \ref{table:IBM_CNOT_gates_opt} for details).

Finally, we examined the consequence of the increased execution time typical for optimisation based synthesis tools on their applicability in practical computational problems.
Although, we can not provide rigorous mathematical statements, our numerical experiments imply that it might be possible to significantly reduce the decomposition time in VQE algorithms by fixing the circuit design.
In Sec.\ref{sec:parametrized}, we provided statistical results for a concrete example to synthesise Pauli exponential circuits parametrised with a single scalar parameter.
The gate synthesis for random parameter values took only $0.82$ seconds in average which is a significant improvement compared to the $\sim10^3$s execution time when the most optimal gate structure is determined.
We believe, that the synthesis time of parametric circuits can be further improved for realistic applications.
We also believe that our results might trigger new strategies to execute useful computing tasks on gate based quantum processors.

\section{Acknowledgements}

The research was supported by the Ministry of Innovation and Technology and the National Research, Development and Innovation Office within the Quantum Information National Laboratory of Hungary, and was
also supported by NKFIH through the Quantum Technology National Excellence Program
(No.2017-1.2.1-NKP-2017-00001) and  Grants No. 2020-2.1.1-ED-2021-00179, K124152, FK135220, KH129601, K134437.
We acknowledge the computational resources provided by the Wigner Scientific Computational Laboratory (WSCLAB) (the formerWigner GPU Laboratory) and the discussions with Christian Gogolin, Gian-Luca Anselmetti, Fotios Gkritsis and Oumarou Oumarou inspiring the synthesis approach discussed in Sec.~\ref{sec:parametrized}.

\bibliographystyle{unsrt}
\bibliography{references}

\end{document}